\documentstyle[11pt,aaspp4,psfig]{article}
 
\def\msun{$M_{\odot}$}
\def\Z{\rm Z}

\def\etal{\it et al. \rm}
\def\feh{\ifmmode {\rm [Fe/H]}\else [Fe/H]\fi}

\begin{document}
 
\title{The Earliest Phases of Galaxy Evolution}
 
\author{Cristina Chiappini}
\affil{Departamento de Astronomia - Observat\'orio Nacional (ON/CNPq), \\
 Cx. Postal 23002, Rio de Janeiro, RJ, Brazil}
\author{Francesca Matteucci}
\affil{Dipartimento di Astronomia, Universit\`a di Trieste, SISSA, \\
Via Beirut 2-4, I-34013 Trieste, Italy}
\author{Timothy C. Beers}
\affil{ Department of Physics and Astronomy\\
Michigan State University, East Lansing, MI 48824}
\and
\author{Ken'ichi Nomoto}
\affil{Department of Astronomy and Research Center for the Early Universe\\
University of Tokyo, Bunkyo-ku, Tokyo 113}
 
 
\begin{abstract}
In this paper we study the very early phases of the evolution of our Galaxy by
means of a chemical evolution model which reproduces most of the observational
constraints in the solar vicinity and in the disk.  We have restricted our
analysis to the solar neighborhood and present the predicted abundances of
several elements (C, N, O, Mg, Si, S, Ca, Fe) 
over an extended range of metallicities
$\feh = -4.0$ to $\feh = 0.0$ compared to previous models.  
We adopted the most recent yield calculations for
massive stars taken from different authors (Woosley \& Weaver 1995
and Thielemann \etal 1996) and compared the results
with a very large sample of data, one of the largest ever used to this purpose.
We have obtained this by selecting the most recent and higher quality
abundance data  from a number of sources and renormalizing 
them to the same solar
abundances.
These data have been analysed with a new
and powerful statistical method which allows us to quantify  the observational
spread in measured elemental abundances and obtain a more meaningful comparison
with the predictions from our chemical evolution model.  
Our analysis shows that the ``plateau'' observed for the [$\alpha$/Fe] ratios
at low metallicities ($-3.0< [Fe/H] <-1.0$) is not perfectly constant
but it shows a slope, especially for oxygen.
This slope is very well reproduced by our model with both sets of yields.
This is not surprising since realistic chemical evolution models, taking
into account in detail stellar lifetimes, never predicted a completely flat
plateau. This is due either to the fact that massive stars of different mass produce 
a slightly different O/Fe ratio or to the often forgotten fact that supernovae of type Ia, 
originating from white dwarfs, start appearing already at a 
galactic age of 30 million years and reach their maximum at 1 Gyr.
For lower metallicities ($-4.0<[Fe/H]<-3.0$) the two sets of adopted yields
differ, especially for iron. In this range the ``plateau'' is almost
constant since at such low metallicities there is almost no
contribution from type Ia supernovae. However, there are not enough data in this domain
to significantly test this point.
Finally, we show the evolution with redshift of the [O/Fe] ratio 
for different cosmologies and conclude that a sharp rise of 
this ratio should be observed at high redshift, {
\it irrespective} of the adopted yields. 
The same behaviour is expected for the [O/Zn] ratio which 
should be easier to compare with the abundances observed in high redshift
Damped Lyman-$\alpha$ systems, as these elements are likely 
not to be affected by dust.
Future measurements of either [$\alpha$/Fe] or [$\alpha$/Zn] ratios in very
metal poor stars will be very useful to infer the nature and the age 
of high-redshift objects.  
\end{abstract}
 

\section{Introduction}
 
The importance of developing an understanding of the earliest phases of galaxy
evolution is clear.   For example, one might hope to explain details of the
heavy element abundance distributions measured in high-redshift galaxies and
the metallic systems associated with damped Lyman-alpha systems ({\it e.g.},
Pettini \etal 1997).  Another important reason is that one can compare the
abundances of extremely-metal-poor stars with the predictions of chemical
evolution models, and thereby impose severe constraints on the theoretical
nucleosynthesis calculations in massive stars ($M \ge 10$ \msun).  The most
massive stars present at the first epoch of star formation were, of course, the
first to restore chemically-enriched gas back into the interstellar medium.
Because of their long main-sequence lifetimes and lack of deep convection
zones, the lower-mass stars ($M \le 0.8$ \msun) which formed at that same time
have preserved the patterns of elemental abundances generated by the initial
burst of star formation.
 
Many studies concerning the predicted nucleosynthetic yields of massive stars
have appeared in the last few years, all of which attempt to take into account
the effect of the initial stellar metallicity on the final yields (Maeder 1992;
Woosley \& Weaver, hereafter WW95; Langer \& Henkel 1996, Thielemann \etal
1996, hereafter TNH96).  The most dramatic differences are found between the
predicted yields from stars with metallicity $\Z=0$ and stars with $\Z \ne 0$,
as shown by the calculations of WW95.  From the standpoint of Galactic chemical
evolution such a difference is not so important, because very few stars with
truly zero metal content must have ever existed, as only a few supernovae (SN)
explosions are required to raise the metallic content of the ISM to a non-zero
value within a few million years.  For stars with $\Z > 0$, the effects of
small differences in initial stellar metallicity on the predicted final yields
of heavy elements appear to be almost negligible.
 
Several studies, such as Arnett (1995), WW95, and TNH96, include approximate
treatments of explosive nucleosynthesis on the final predicted yields of
chemical elements.  The effects of mass loss by stellar winds in very massive
stars ($M > 30-40$ \msun ) on the nucleosynthetic yields has been investigated
by Maeder (1992), Woosley \etal (1993; 1995), and Langer \& Henkel (1996).  The
net effect of mass loss is to decrease the total yield of metals produced by
massive stars because of the helium which is subtracted from further
processing. We took into account the presence of mass loss in massive 
stars by adopting the mass-mass of the He-core relation of Maeder and Meynet (1989).

There are, of course, many uncertainties still present in modern
nucleosynthesis calculations, as is aptly illustrated by the disparity among
various studies concerning the yield of iron (and other iron-peak elements)
produced in massive stars.  In the WW95 study, for example, the iron yield does
not exhibit a monotonic behaviour as a function of the initial stellar mass, as
it {\it decreases} for masses larger than $\simeq 35$ \msun.  By contrast, in
TNH96 the iron yield is an ever-increasing function of the initial stellar
mass, but it is systematically lower than the yields of WW95 for masses smaller
than $\simeq$ 30 \msun .  The reason for these differences can be traced to the
different physical inputs to the two sets of calculations, namely, the iron and
helium core masses, the assumed rate of the $^{12}C(\alpha,\gamma)^{16}O$
reaction, differences in the adopted explosion mechanism, etc. (see WW95 for a
detailed discussion).
 
Given the ``flexibility'' of the predicted yields of heavy elements, it is of
interest to explore the nature and the magnitude of differences which arise in
models of Galactic chemical evolution as a result of the use of the different
yields.  In this paper, we compare yields from different authors by means of a
model of chemical evolution for our Galaxy which reproduces the main
observational features (Chiappini \etal 1997), and compare the predictions with
one of the largest sample of abundances ever used for this purpose (see also Samland 1998).   
Particular emphasis is given to the predictions
for very-metal-poor stars since they place the strongest constraints on the
available yields for massive stars and on the different
theories for the progenitors of type Ia SNe.

In performing this comparison we
use a statistical analysis discussed by Ryan \etal (1996) in order to define
the best fit to the data, which always show a considerable spread, and to
compare these best fits with the theoretical predictions.  
The main aim of this paper is to provide the best fits of a large number 
of elemental ratios as functions of [Fe/H] and compare them with detailed 
model predictions in order 
to impose constraints on the nucleosynthesis in massive stars, type Ia SN progenitors, 
timescale for the formation of the halo and thick disk and finally to predict the behaviour
of the relative abundance ratios as functions of redshift in
the context of different cosmological models. 
In fact, abundance ratios can be used either as cosmic clocks
or to infer the nature of high -redshift objects.
 
The paper is organized as follows.  In \S 2 we summarize the available
observational data from literature studies of the elemental abundances in
metal-poor stars. In \S 3 we describe the statistical method adopted to
describe the best fit to the observed data.  A brief description of the
chemical evolution model is given in \S 4.  In \S 5 we discuss a prediction of
the [O/Fe] ratio for objects formed at high redshift.  In \S 6 our results are
presented and some conclusions drawn.
 
\section{Observational Data}
 
Over the past few decades, many studies of very-metal-poor stars have been
carried out ({\it e.g.}, Wallerstein \etal 1963; Spite \& Spite 1978; Luck \&
Bond 1985; Magain 1987; Gilroy \etal 1988; Peterson \etal 1990; Gratton \&
Sneden 1988,1991; McWilliam \etal 1995; Ryan \etal 1991; 1996).  The most
recent results, based on stars selected from the HK survey of Beers \etal
(1985; 1992), have probed the nature of the relative abundances of elements in
large samples of stars at extremely low metallicity.  In the McWilliam \etal
(1995) study, for instance, the abundances of several heavy elements are
derived for 33 extremely-metal-deficient giants, with $-4.0 \le \feh \le -2.0$.
They found previously unnoticed trends of [Cr/Fe], [Mn/Fe] and [Co/Fe] with
\feh . Both [Cr/Fe] and [Mn/Fe] exhibit a decline of about 0.5 dex with
decreasing iron abundance between $\feh = -2.4$ and $\feh = -4.0$. On the other
hand, [Co/Fe] increases by about 0.5 dex over the same range in iron abundance.
They confirmed the well known decline in [Al/Fe], [Sr/Fe] and [Ba/Fe] with
decreasing [Fe/H]. The remarkable fact is that all six of these abundance
ratios show a sudden change of slope near $\feh=-2.4$.  McWilliam
\etal concluded that a distinct phase of nucleosynthesis occurred before the
Galaxy reached an abundance $\feh \sim -2.4$, their preferred scenario being
that of variable stellar yields, either changing with mass, metallicity, or
both.
 
Concerning the $\alpha$-elements, McWilliam \etal found [Mg/Fe], [Si/Fe] and
[Ca/Fe] to be overabundant relative to iron ([$\alpha$/Fe] $= +0.44 \pm 0.02$
dex), confirming the trends found for other halo stars at higher abundances.
The element titanium, often considered to be an alpha element, exhibited a
smaller overabundance relative to iron, [Ti/Fe]$=+0.31$ dex.  This difference
was interpreted as due to a larger contribution to Ti from type Ia SN.
However, at very low metallicities [Mg/Fe] appears to decrease whereas [Ti/Mg]
and [Ca/Mg] increase by $\simeq$0.2 dex from $\feh=-2.0$ to $\feh = -4.0$.
McWilliam \etal interpreted this as the possible effect of
metallicity-dependent yields.  Another remarkable result of their study is the
large spread (up to 3 dex for a given \feh) observed in the abundance ratios of
some of the heavy elements, especially in [Sr/Fe] and [Ba/Fe], which they
interpreted as intrinsic and due to incomplete mixing of the ISM of the early
Galaxy.  The same conclusion had been already reached by Ryan \etal (1991) and
Primas \etal (1994) from the spread they observed in [Sr/Fe] and [Ba/Fe] ratios
at very low metallicities.
 
Ryan \etal (1996) obtained abundances for 19 extremely-metal-poor stars (giants
{\it and} dwarfs). They found that the $\alpha$ elements (Mg, Si, Ca and Ti)
have almost uniform overabundances relative to iron down to [Fe/H]$\simeq -$4,
which is the current limit of observations.  They also observed a slight
increase in the [Mg/Fe] ratio at $\feh < - $2.5. These authors found that Ti
behaves like the other $\alpha$-elements and confirmed the underabundance of
[Cr/Fe] and [Mn/Fe], and the underabundance of [Co/Fe], in stars with $\feh <
-2.5$, as reported by McWilliam \etal .  These authors concluded that there is
a real star-to-star abundance differences at the lowest metallicities, and that
this spread exists for both dwarfs and giants, indicating that the observed
spread cannot be explained by appealing to mixing processes within the stars
themselves. Ryan \etal also discussed a model in which an assumed correlation
between SN iron-peak yields and explosive energies is able to explain the
existence of well-defined ({\it low} scatter) trends among a number of heavy
elements (such as Cr, Mn, and Co) down to an abundance $\feh \sim -2.7$,
despite the fact that enrichment of the ISM of the proto-halo took place under
chaotic and inhomogeneous mixing conditions. 
 
The most important recent study on disk stars is that of Edvardsson \etal
(1993), who derived abundances of O, Na, Mg, Al, Si, Ca, Ti, Fe, Ni, Y, Zr, Ba
and Nd as well as individual photometric ages for 189 nearby disk population F
and G dwarfs. They also estimated, from kinematical data, the orbital
parameters of the stars  and derived estimates of the Galactocentric distances
of the stellar birthplaces.  From their high signal-to-noise spectra and a
careful treatment of stellar temperatures, Edvardsson \etal  obtained iron
abundances and abundance ratios relative to iron for most elements with
accuracies at a level of 0.05 dex.  They found a considerable variation in the
metallicity of stars formed at a given time in the disk, implying only a weak
correlation between age and metallicity. Their result strongly suggests there
is no single age-metallicity relation, but rather {\it several} such relations
due to different chemical evolution rates in different parts of the Galaxy.
Metal-poor ($\feh < -0.4$) stars in their sample were shown to be relatively
overabundant in the $\alpha$-elements, confirming all the previous studies, but
the most interesting finding is that the [$\alpha$/Fe] ratios for these
metal-poor stars {\it decreases} with increasing Galactocentric distance.  This
implies that star formation started first and proceeded more vigorously in the
Galactic center than farther out in the disk, confirming theoretical
suggestions about an inside-out formation of the Galaxy (Chiosi \& Matteucci,
1980; Matteucci \& Fran\c cois 1989, hereafter MF89, Burkert \etal 1992).  The
scatter found in abundance ratios such as [Si/Fe] is only about 0.05 dex, about
50 times less than the corresponding range in [Fe/H].  In their work they
discussed possible explanations for this effect, including orbital diffusion
and infall of unenriched material. On the other hand, Fran\c cois \& Matteucci
(1993) showed that orbital diffusion, coupled with the fact that the Galaxy
formed at different rates at different Galactocentric distances, can explain
most of the observed spread in \feh\ and [$\alpha$/Fe], and also the fact that
the spread in the abundance ratios is smaller than the range in absolute
abundances.
 
In this paper we will restrict our attention only to C, N, O, Mg, Si, S, Ca,
and Fe as they are the elements for which data of good quality are available
and because they allow us to study in particular the [$\alpha$/Fe] ratios,
which are very useful tools to constrain stellar nucleosynthesis, Galactic
evolution, and the nature of high-redshift objects.
 
\subsection{Oxygen}
 
The element oxygen deserves a special discussion.  The main concern in
abundance studies of this element arises due to the paucity of samples of
significant size analysed in a homogeneous way.  In combined samples, the
possible existence of systematic offsets between different sets of data and the
discrepancy between O abundances determined using high-excitation permitted and
low-excitation forbidden lines (the first usually observed in dwarfs, the
latter in giants) limits our ability to interpret observed trends.  In an
effort to remove the discrepancy between dwarfs and giants,
Gratton \etal  (1996) reanalyzed the abundances of O and Fe in a large sample
of stars in the solar neighbourhood, taking into account recent higher
temperature estimates in the analysis of dwarfs, and departures from the LTE
assumption when considering the formation of high-excitation permitted OI
lines.  They found that the [O/Fe] ratio is nearly constant ([O/Fe] $\simeq
+$0.5 dex) among inner-halo and thick-disk stars.  Moreover, their data
suggest that the [O/Fe] decreases by $\simeq$ 0.2 dex while the [O/H] ratio
remains constant during the transition between thick- and thin- disk phases,
indicating a sudden decrease in star formation at that epoch.

In a recent paper Israelian et al. (1998) performed a detailed
abundance analysis of 23 unevolved metal-poor stars and found that the [O/Fe]
ratio increases from 0.6 to 1.0 between [Fe/H] -1.5 and -3. 
In that paper the abundances were determined using the OH bands in
the near UV, with high resolution. They also address the long standing
controversy, mentioned above, between oxygen abundances from forbidden and permitted 
lines and show that this can be resolved. 
As our paper
was already submitted these data were not included in the present investigation
but their implications are discussed in \S 3.
 
\subsection{Carbon}
 
The results on carbon are still somewhat controversial.  Laird (1985)
determined carbon abundances in a large sample of halo an disk dwarfs, and
found [C/Fe]$ = -0.22$ dex , with an intrinsic scatter of 0.1 dex, over the
entire metallicity range considered ($-2.5 \le \feh \le +0.5$).  However, an
analysis of the [C/Fe] ratio as a function of the effective temperature
indicated a systematic offset in the derived abundances; a correction of 0.20
dex was applied to all the [C/Fe] determinations, resulting in roughly solar
values.  Tomkin \etal (1986) examined 32 halo dwarfs with abundances in the
range $-2.5 \le \feh \le -0.7$ dex and found that [C/Fe] $=-0.2 \pm 0.15$ dex
for $\feh >-1.8$ dex.  Carbon \etal (1987) derived carbon abundances for 83
dwarfs in the metallicity range $-2.5 \le \feh \le -0.6$.   They found a solar
[C/Fe] ratio over most of the metallicity range, with a star-to-star scatter of
0.18 dex. They also noticed a slight increase of [C/Fe] at very low
metallicities, but noted that this upturn is sensitive to the assumed O
abundance.  Wheeler \etal (1989) reanalyzed all three surveys and attempted to
place them onto a common effective temperature scale.  The overall trend of
slightly increasing [C/Fe] ratio at low metallicities did not disappear.

One important recent result which may be of considerable significance to the
chemical evolution of the early Galaxy is the fact that the {\it most}
metal-poor stars exhibit a surprisingly high incidence of carbon enrichment.
Roughly 15\% of the stars in the sample of Beers \etal (1992) with $\feh <
-2.5$ have CH (G-band) features which are stronger than normal (see figure 1 of
Norris, Ryan, \& Beers 1997a).  Large [C/Fe] ratios appear to exist among a
number of extremely-metal-poor dwarfs and subgiants, as well as giants.  Among
the carbon-rich stars at extremely low abundance is the r-process enhanced
star CS 22892-052, discussed in detail by Sneden \etal (1994, 1996), and Norris
\etal (1997a).  High-resolution spectroscopic studies of other stars with $\feh
< -2.5$ and large carbon abundances have been presented by Barbuy \etal (1997),
Norris \etal (1997a), Norris \etal (1997b), and Bonifacio \etal\ (1998).  The
most extreme cases exhibit [C/Fe] ratios roughly 2 dex above the solar value.  
The mechanism by which carbon has been produced and/or enhanced in these
extreme stars is still unknown.  Since it remains possible that all or most of
the carbon-enhancement may be due to mixing processes within the stars
themselves, or due to the transfer of material from an evolved companion star,
we have chosen not to include these stars in the data for our present study. 

The apparently constant and solar [C/Fe] ratio over the range from $\feh \sim
-2.5$ to disk metallicities would seem to indicate that C and Fe, although
produced through different nucleosynthesis mechanisms (C being produced through
the triple$-\alpha$ process during hydrostatic helium burning in stars of all masses,
whereas Fe is synthesized in explosive burning conditions), are originating
from similar stellar mass ranges.  In other words, the two elements should be
mainly produced in the same relative proportions.  Anderson \& Edvardsson
(1994) determined the [C/Fe] ratio in 85 dwarfs in the metallicity range $-1.0
\le \feh \le +0.25$ and found that this ratio is slowly decreasing with time
and increasing metallicity in the disk.  This would indicate a slightly more
significant production of carbon from massive stars relative to their iron
production.  Moreover, Tomkin \etal (1995) determined C abundances in 105
dwarfs in the range $-0.8 \le \feh \le +0.2$ and found a moderate enrichment of
C in metal-deficient stars, on the order of [C/Fe]$=+0.2 \pm 0.05 $ at
$\feh=-0.8$.  Such a behaviour is qualitatively similar to that seen for
$\alpha$-elements and oxygen.  The carbon isotope $^{13}$C seems to show a
predominantly secondary origin, confirmed by the behavior of the
$^{12}$C/$^{13}$C ratio in the ISM along the Galactic disk (increasing with
decreasing iron abundance --  see Wilson \& Matteucci, 1992; Wilson \& Rood
1994).
 
\subsection{Nitrogen}
 
Nitrogen abundances derived from stellar spectroscopy, especially at low
metallicities, are the most uncertain.  Laird (1985) determined nitrogen
abundances in 116 dwarfs.  The survey covered the metallicity range $-2.5
\le \feh \le +0.5$, with most of the stars having $\feh > -2.0$.
He found [N/Fe]$=-0.67 \pm 0.2$ over the entire metallicity range. 
As discussed by Laird (1985) the reason for this offset is not clear and
may be related to a systematic error in the temperature scale. He then
applied an offset of $+$0.65 dex to all the [N/Fe] ratios.
 
Carbon \etal (1987) derived nitrogen abundances for 83 dwarfs in the range
$-2.5 \le \feh \le -0.6$, with the majority of stars having $\feh < -1.5$. 
In the present investigation we are adopting the Carbon et al. data for N
without any correction for a Teff dependence (see Carbon et al. for a 
discussion on this correction). From figure 2 it can be seen that when
the raw data from Carbon et al. are adopted they are not inconsistent
with a secondary N scenario for massive stars although any conclusion
about this point still waits for better data. If, as suggested by
Carbon et al., their data should be corrected for this temperature
dependence, then [N/Fe] turns out to be roughly constant over the entire
metallicity range. In that case, as for carbon, this implies that both Fe and N,
although they have a different nucleosynthetic origin, are produced in the same
proportions in the same stellar mass ranges.  
However, as stressed by Wheeler et al. (1989) we still are left with considerable
uncertainty in the preferred [N/Fe] value even if it is constant with metallicity.

The isotope $^{15}$N in the ISM
has been recently studied by Dahmen \etal (1995), who derived a Galactic
gradient of [$^{14}$N/$^{15}$N] of 19 $\pm$ 8.9 dex kpc$^{-1}$, and concluded
that $^{15}$N should be considered a secondary element (normally ``secondary''
means produced proportionally to the initial stellar metal content, although
the same behaviour can be expected if an element is produced without a
metallicity dependence but restored on long timescales), and that $^{14}$N has
a significant primary component.

\section{Observed Trends and Scatter}
 
We have made an effort to obtain a more quantitative comparison between model
predictions and observations of the elemental abundance ratios elements as a
function of metallicity.  Accordingly, we include in our data set essentially
all published observations of the elements under consideration (C, N, O, Mg,
Si, S, Ca and Fe), and renormalize all data to a common solar value (the Anders
\& Grevesse 1989, hereafter AG89, meteoritic values). This is especially
important for the adopted solar iron abundance, which has changed (in the
literature) by about 0.2 dex in the last 10 years.  We recognize, however, that
this normalization alone is not enough to obtain an homogeneous sample and
that, for instance, we should also correct for the different adopted oscillator
strengths ({\it gf}).  In the case of a differential analysis the correction
for the {\it gfs} should not be very important whereas for the absolute
analysis this correction may not be not negligible. However, this effect is not
as important for metal-poor stars as it is for more metal-rich ones (Fran\c
cois private communication).
 
As the amount of data for the relative abundances of elements in
samples of metal-poor stars increases, it is important to consider improvements
in the methodology which we use to describe and analyse the trends which are
revealed.  Following the approach described by Ryan \etal (1996), we have made
use of the summary statistics of Cleveland \& Kleiner (1975).  After applying
corrections to the literature values of abundance ratios (as described above)
we obtained a straight average for each elemental ratio.  We then obtain three
lines which summarize the trend of the data as a function of \feh.  The three
lines are:  MM -- the midmean -- an average of all data between the
quartiles of the distribution of relative abundances over a given range in
\feh; LSMM -- the lower semi-midmean -- the midmean of all observations below
the median; USMM -- the midmean of all observations above the median.  If the
data are scattered about the midmean according to a normal distribution, the
semi-midmeans are estimates of the true quartiles.  We have checked this
assumption, and it appears to hold, at least globally (over all values of
\feh) for each of the abundance ratios considered herein.  In Figures 1 and 2
we show plots of the resulting summary lines, obtained by taking either 10 or
15 data points in the estimation window, depending on the density of data.  The
final plots represent locally weighted regression ({\it lowess}; see Cleveland
\& Devlin 1988) smooths of the summary lines.
 
Figures 1 and 2 show our plots for O, Mg, Ca, Si, S, C and N respectively.
Those elements were chosen not only because they are the most abundant and
consequently less uncertain but also because for even-z elements the yields are
less dependent on the metallicity (see WW95 and Samland 1997). Samland shows
that the stellar models tend to underestimate the yields of odd-z elements, but
that there is a good agreement for the even-z ones.  The left panels of these
figures show the original data sets quoted in Table 1 compared to the models; 
the right panels show a comparison between the trend lines obtained from the
data and the model predictions (see a description of the different models in
the next section).

At a first glance of figures 1 and 2 one sees that the so-called plateau for the 
[$\alpha$/Fe] ratios at low metallicities is not really constant.
This is an important fact and it is very likely to be real.
In fact, it is not surprising that the plateau is not constant, since 
theoretical models always predict a slight decrease of [$\alpha$/Fe] ratios during
the halo phase and this is due to the fact that massive stars of different masses 
produce slightly different O/Fe ratios. 
In addition, as predicted by the canonical model for SNe Ia 
(Matteucci and Greggio, 1986) the first SNe appear already after 30 million years 
although they reach a maximum at around 1 Gyr from the beginning of the evolution.
This epoch corresponds to a metallicity of [Fe/H] $\sim$-1.0
and to the more drastic change of slope in the [$\alpha$/Fe] ratios.
Only simple models where the SN Ia are imposed to appear only after 1 Gyr 
produce a flat plateau.
This large new collection of data is therefore important since it shows 
clearly this effect.

Recently, Israelian et al. (1998) presented new [O/Fe] data, not included in this study, 
showing that oxygen is overabundant relative to iron, with the [O/Fe] ratio
increasing from +0.6 to +1.0 going from [Fe/H]=$-$1.5 to $-$3.0. This overabundance
is larger than it is shown by the data used in this paper. Moreover they
show that the plateau of the [O/Fe] ratio is not constant but it has a slope
of $-$0.31 $\pm$ 0.11 in the range $-$3.0$<$[Fe/H]$<-$1.0. 
 
It is instructive to consider the spread of the abundance ratios about the
underlying trend lines over several intervals of \feh.  To do this, we first
obtain the {\it lowess} line for each elemental ratio as a function of
\feh, then derive the residual in the ordinate of each data point about the
trend.  In Table 2 we summarize robust estimators of scale for these residuals
over the metallicity intervals:  $-0.5 > \feh $; $-1.5 > \feh \le -0.5$;
$-2.5 < \feh \le -1.5$; $\feh \le -2.5$.  The scale estimator we employ, the
biweight estimator $S_{BI}$ discussed by Beers \etal (1990), has been shown to
have the desirable property that it is not overly influenced by outliers, and
approaches the more-commonly-applied standard deviation estimator of scale for
samples drawn from a normal parent distribution. For a number of elements we
had to combine stars across several abundance intervals due to the paucity of
extant data --  entries in the table span the columns over which the abundances
have been combined.  For each entry in the table we also include an estimate of
the 68\% (one-sigma) bootstrap confidence interval about the biweight estimate
of scale, indicated as $IS_{68}$, which is useful for assessing the
significance of the change in spread from one abundance interval to the next.
In the table we also indicate an estimate of the 95\% range in the residuals,
obtained as $R_{95} = 2\; {\rm x}\; 1.96\; S_{BI}$, since 1.96 $S_{BI}$ is the
one-sided two-sigma interval of a normal distribution.
 
Figure 3 summarizes the data shown in Table 2. This figure shows that the
scatter in the abundance ratios {\it increases} with decreasing metal
abundance, at least for Ca, Mg, and Si (the alpha elements for which
substantial amounts of data are available).  For O the scatter is essentially
constant over the metallicity range but the statistics are poor, although
the sample is a more homogeneous one.  
In the case of N the data show an
increase in the scatter of the [N/Fe] ratio as metallicity decreases. However,
as already discussed in \S 2.3 the data for low metallicities are
very uncertain and are probably affected by the [N/Fe] vs Teff correlation
found by Carbon et al. (1987). The trend lines in Figure 2 show an
obvious decrease in [N/Fe] at low metallicity (when the raw data of Carbon
et al. are adopted), but no data presently exist for stars with $\feh \le -3.0$, 
so we cannot be certain not only that this trend exists but even if it continues to
lower abundances.
The scatter in the [C/Fe] ratios also increases as the
metallicity decreases.  For this element the trend lines in Figure 2 show a
small bump around $\feh = -0.5$ and a slight increase in the [C/Fe] ratio for
metallicities $\feh < -2$ (already noted by Carbon et al. 1987).

\section{The Chemical Evolution Model}
 
The model of chemical evolution we adopt here is that of Chiappini \etal
(1997), where a detailed description can be found.  We recall the main
assumptions of this model:
 
\medskip
\begin{itemize}
\item 
The Galactic disk is approximated by several independent rings, 2 kpc wide,
without exchange of matter between them.
Continuous infall of gas ensures the
temporal increase of the surface mass density $\sigma_{Tot}$ in each ring.
 
\item 
The instantaneous recycling approximation is relaxed.  This is of fundamental
importance in treating those isotopes, such as $^{14}$N and $^{56}$Fe, which
are mostly produced by long-lived stars.
 
\item 
The prescription for the star formation rate (SFR) is:
$$
SFR \propto \sigma_{Tot}^{k_{2}} \sigma_g^{k_{1}} \eqno (1)
$$
 
\noindent where $\sigma_{Tot}$ is the total surface mass density and $\sigma_g$
is the surface gas density and $k_{1}=1.5$ and $k_{2}=0.5$.  A threshold in the
surface gas density ($\sim 7~ M_{\odot}pc^{-2}$) is also assumed; when the gas
density drops below this threshold the star formation stops. The existence of
such a threshold has been suggested by star formation studies (Kennicutt 1989).
 
\item 
For the initial mass function (IMF) we adopt the prescriptions 
from Scalo (1986) as described in Chiappini \etal (1997).
 
\item 
This model assumes that the halo+thick disk and the thin disk are formed during
two different infall episodes--the thin disk does not form out of gas shed from
the halo and the thick disk, but simply out of external gas.  Such an
interpretation is supported by recent dynamical and kinematical studies of
stars in the outer Galactic halo by Sommer-Larsen \etal (1997).
Under these
hypotheses, the infall rate is given by:
 
 $$ {dG_i(r,t) \over dt}={A(r)(X_{inf})_{i} e^{-t/\tau_T} \over
\sigma_{Tot}(r,t_{G})} + {B(r)(X_{inf})_{i} e^{-(t-t_{max})/\tau_D(r)} \over
\sigma_{Tot}(r,t_{G})} \eqno (2) $$
 
\noindent where $\tau_T$ represents the time scale for the formation of the
halo and the thick disk, and $\tau_D(r)$ represents the time scale for disk
formation, which is assumed to increase with Galactocentric distance.  The
best-fit model of Chiappini \etal (1997) suggests that the timescale for the
formation of the halo and thick disk is quite short ($\sim 0.5-1.0$ Gyr)
whereas the timescale for the formation of the thin disk is quite long ($\sim
8.0$ Gyr for the solar vicinity). This timescale for the thin disk ensures a very good fit of the new
data on the G-dwarf metallicity distribution (Wyse \& Gilmore 1995; Rocha-Pinto
\& Maciel 1996).  A(r) and B(r) are derived by the condition of reproducing
the present total surface mass density distribution in the solar vicinity.
$(X_{inf})_{i}$ is the abundance of the element $i$ in the infalling
material and $T_G$ the age of the Galaxy.
\end{itemize}
 
The contributions to the chemical enrichment from supernovae of different type
(Ia, Ib and II) as well as from stars dying as C-O white dwarfs and
contributing processed elements through stellar winds and the planetary nebula
phase are taken into account in great detail (see Chiappini \etal 1997).
 
The adopted nucleosynthesis prescriptions are from: (a) Renzini \& Voli (1981)
for intermediate stellar masses and (b) Thielemann \etal (1993) and
Nomoto \etal (1997b) for SNe Ia. For
the massive stars we considered  two different sets of yields, one from WW95
and the other from TNH96.
 
Figures 4 and 5 show comparisons between both sets of yields for massive stars.
In these figures we show also the abundance yields adopted by MF89. WW95
present the yield calculations for stars of up to 40 \msun, whereas TNH96
obtained the yields for stars of up to 70 \msun (where the yields for the 18, 
40 and 70 \msun stars are given in Nomoto \etal 1997a and Tsujimoto \etal 1995).  
Both sets were extrapolated
up to $M_{up}$ = 100 \msun, though this has little effect on the results for
the chemical evolution of the Galaxy, as the adopted IMF predicts very few
stars more massive than 40 \msun.  Matteucci (1996) shows a comparison of
yields relative to stars of different metallicities for some of the most
important heavy elements.  She shows that there are non-negligible differences
in the yields from $Z=0$ and $\Z \not= 0$ in the WW95 calculations (at least
for C, O, Mg, Si and Fe) whereas the yields for different initial metallicities
greater than zero are very similar ($\Z$ = 0.0001, 0.0002, 0.0020 and 0.0200).
As discussed in the introduction, from the point of view of Galactic evolution
this difference is not very important since very few stars of zero metallicity
must have existed.  WW95 present several tables for different metallicities,
with the ejected mass as a function of initial stellar mass.  We adopt the
table which corresponds to the calculation made for $\Z = \Z_{\odot}$.
 
The main differences seen in Figure 5, for the elements Si, Ca and S, are due
to the adopted yields in MF89 and the present ones.  The difference originates
in the fact that both the sets of recent yields contain explosive
nucleosynthesis calculations.  The yields for massive stars adopted by
MF89 were from Woosley \& Weaver (1986), which did not include the explosion
mechanism in their calculations.  To overcome this problem MF89 assumed that
the fraction of Si-Ca elements which fall back into the collapsing cores of
the post-explosion remnants were roughly 40\% of what is remaining of the Si-Ca
elements after subtracting the iron.  As discussed by these authors, this value
was uncertain and was chosen to give the best agreement with the
observations.  The abundances of elements lighter than silicon are almost
unmodified by the explosion (see WW95) and the nuclei which are ejected are the
same ones present in the pre-supernova star.  The yields for S, Si and Ca
obtained by TNH96 are systematically larger than those obtained by WW95 for
stars more massive than 30 \msun , whereas they are smaller for masses below
this value. The elements Si, S and Ca are products of explosive
oxygen burning, and the difference of their yields may stem from (the difference in the explosion energy and/or) the difference in the presupernova
density structure of the innermost oxygen layer due to the different
treatment of convection.

The yields of the most abundant isotopes are not very different. The 
first example is $^{16}$O, which is the most abundant heavy element
made in massive stars. Oxygen is produced by helium-burning and neon-burning,
and its synthesis is expected to be roughly independent of metallicity. The main
source of uncertainty for the $^{16}$O and $^{12}$C yields is ascribed to the
uncertainties in the $^{12}$C($\alpha,\gamma$)$^{16}$O reaction, although the
$^{12}$C production is also sensitive to details of convection (see WW95 for a
discussion).
 
The iron yield from WW95 is higher than the TNH96 value for stars with masses
below $\simeq$ 35 \msun . 
The
iron (and iron-group elements) yield depends strongly on the 
mass cut that divides the ejecta and the material fallen back
onto the neutron star or black hole, thus being sensitive to the explosion
mechanism. Hence the predicted yields are uncertain by (at least) a factor of
two. In fact, Timmes \etal (1995) suggested that the iron yields from WW95
should be lowered by just this factor. This would lead to an agreement with the
TNH96 values for these elements.
 
For the element Mg we can see a non-negligible difference between both
sets of calculations, which would primarily stem from the 
different treatment of convection (i.e., stability criterion and 
semi-convective mixing). The WW95 values are much smaller than the TNH96 ones.
Chiappini \etal (1997) adopted in their chemical evolution model the WW95
yields, and suggested that it might be appropriate to  increase the Mg yields
in order to obtain a solar value of Mg in agreement with the observed one.
 
In Figure 6 we present the masses ejected in the form of various elements,
weighted by the IMF in the range of massive stars, as a function of the initial
mass of the progenitor star for WW95 (upper panel) and TNH96 (lower panel) and
calculations, respectively.
 
The new calculations also include the yields for $^{14}$N.  Figure 7 presents
the same plot as Figure 6 for this element (solid line: WW95 and dot-dashed
line: TNH96). $^{14}$N in massive stars is thought to be mostly secondary
although some primary contribution can occur.
 
In Table 3 we show the predicted and observed solar abundances of the elements
under consideration adopting each of the two sets of predicted yields (model A
with WW95 yields and Model B with the TNH96 yields) and the solar abundance
values from Anders and Grevesse (1989).  
The differences between columns (2) and (3) reflect the
differences in the adopted yields for massive stars (in Table 3, the exponents
are shown in parentheses).  We can see that the difference for $^{14}$N in the
final abundances is small, even with the very different predictions seen in
Figure 7, as this element is mainly produced by stars with progenitors in the
intermediate-mass range. The yields adopted for the intermediate-mass range
are the same in both models (Renzini \& Voli 1981).  Note the good
agreement for the solar iron abundance when the TNH96 values are adopted.  In
\S 6 we discuss the differences in the predicted abundance
ratios as a function of metallicity for the two sets of yields considered in
this paper.

\section{The Abundance Ratio of [O/Fe] at High Redshift}

In figures 8 and 9 we show the predicted [O/Fe] ratios as functions of
redshift for different cosmologies. In particular, a cosmology is identified by
the age of the Universe $T_U$, the redshift of the galaxy formation $z_f$, the
density parameter $\Omega_0$ and the cosmological constant, $\Lambda$.

Figure 8 shows the predictions of the model which adopts the yields of WW95;
Figure 9 shows the predictions using the yields of THN96.  A striking
characteristic in both figures is the sharp increase of [O/Fe] at very high
redshift. The spike in [O/Fe], which is independent of the adopted yields, is
the result of the  rapid increase of metallicity at the initiation of the star
formation process.  For instance, for the cosmological model with $T_U$=15
Gyrs, $z_f$=5, $\Omega_0$=1 and $\Lambda$=0 (solid line in figures 8 and 9),
the initial metallicities are $\feh = -5.8$ and $-5.4$ 
for TNH96 and WW95,
respectively. 
As already mentioned, it is sufficient that some Type II SNe
explode to raise the metal content of the interstellar medium from zero to
metallicities of the order of [Fe/H] $\sim -5$, already in the first 0.5 Gyrs.
For the solid line model, in figure 8, we have that $\feh = -5.4$ at a redshift
of 3.52. The [O/Fe] plateau extends until a metallicity of about $\feh = -3.0$,
which corresponds to a redshift value of 3.51 (in other words, a range
in metallicity of $\Delta \feh = -2.4$ corresponds to a change in redshift
of $\Delta$z = 0.01). This explains the peaks when plotting the abundance
ratios as functions of redshift instead of metallicity. A similar result
applies when adopting the TNH96 yields.

The same behavior is expected for the [O/Zn] ratio which should be easier to compare
with the abundances observed in high redshift DLA systems, as these elements are likely
not to be affected by dust, and because the abundance of Zn has been
measured for a large number of such systems (Pettini et al. 1997).
The element Zn behaves exactly like iron and therefore is very probably
formed in the same way, namely mostly during the explosion of a type Ia SN
(see Matteucci et al. 1993).

This striking feature can be used to identify proto-spiral galaxies
similar to our own Galaxy when observing high redshift objects. This is
an important point in view of the fact that more and more data are
becoming available for damped Lyman-$\alpha$ systems at high redshifts. 

\section{Results and Conclusions}
In this paper we computed the chemical evolution of the solar neighbourhood
by adopting two different sets of yields for massive stars with the aim
of imposing constraints on the early phases of galaxy evolution.
To do that we have assembled one of the largest sample of  
relative abundance ratios observed in a range of 
metallicity going from [Fe/H]=$-$4.0
up to solar.
We have adopted a detailed method
to compare model predictions and observations in a self-consistent way.
This method allowed us to quantify the apparent spread in the data,
to discover some new trends and to impose constraints on stellar nucleosynthesis.\par
Our main conclusions can be summarized as follows:
\par
- From the analysis of the data sample we found that the so-called 
``plateau'' for the [$\alpha$/Fe] ratio at 
low metallicities is not perfectly flat but it presents a slight slope, 
especially for oxygen. This is an important fact that had been not 
noticed before since the number of data at low metallicities 
used to compare with chemical evolution models was always small 
(except in Samland 1997).
This slight slope is in very good agreement with our model predictions 
for both sets of yields in the metallicity 
range ($-$3.0$<$ [Fe/H] $< -$1.0) 
and it is due
partly to the fact that massive stars of different mass produce slightly 
different O/Fe ratios 
and partly to the appearence of SNeIa which start occurring
already after 30 million years in the adopted model of a white dwarf 
plus a red giant companion (Matteucci and Greggio 1986). These 
supernovae reach then a maximum at around 1Gyr which is the time 
at which the [O/Fe] ratio 
starts changing more drastically.
This corresponds, in our model, to  roughly a metallicity of [Fe/H] = $-$1.0.
If we identify this time with the end of the halo phase, although the border line between halo and thick-disk is not so precise, we can infer that 1 Gyr is the average timescale for the formation of the halo.
\par
- By comparing the different sets of yields used in this paper we can not draw very firm conclusions since both of them can reproduce the solar abundances inside a factor of two. However, going more into detail we can say
that the model predictions are in agreement
with the observed values, except for two elements which are only marginally
consistent with observations, namely, Mg when the WW95 yields are adopted, 
and
$^{12}$C when the TNH96 yields are adopted. 
Moreover, the values of Mg and $^{12}$C
are always underestimated either adopting TNH96 or WW95 yields.
For Mg this has been already shown by  
Chiappini et al. (1997) and Thomas et al. (1997). In the case of $^{12}$C
the TNH96 yields are systematically lower than the WW95 ones, and this could
be the origin of the ``valley'' seen in figure 2 for [C/Fe] between the 
metallicities [Fe/H] = $-$1 and $-$3.
For oxygen both sets of yields are very similar, at least up to 30
\msun . However, for masses above 30 \msun\ the yields obtained by TNH96 are
larger. While the TNH96 yields for iron in massive
stars increase almost linearly with mass up to 40 \msun\, being
roughly constant after this value, the calculations by WW95 indicate
an iron production which is concentrated in the range between 
20 and 30 \msun .  

The
predicted iron solar abundance obtained when adopting the 
TNH96 yields is closer
to the observed value although the difference between the TNH96 and WW95 values
are smaller than the uncertainties involved. 
It can be seen that the O/Fe ratio for very high mass stars varies
with the stellar mass
much more in WW95 predictions than in the TNH96 ones. This influences  
the slope of the [O/Fe] plateau at low metallicities and can be used as a
constraint to the yields of massive stars. Unfortunately, at such low
metallicities ([Fe/H] between $-$4 and $-$3) there are still not enough data to
be able to choose between the two set of yields.
For Si, S, and Ca the WW95 yields are higher than the TNH96 ones up to 35
\msun . The opposite occurs for masses above 35 \msun .  In this case
both models predict [X/Fe] ratios in good agreement with the 
observational data.
This happens because the differences between predicted iron
and $\alpha$-element yields are compensating each others. 

-The observed increasing scatter in the abundance ratios of many elements with
decreasing metallicity could be interpreted as if the interstellar medium was
less homogeneous at early times. It seems unlikely that the increasing scatter
is due solely to the increased uncertainty in the
estimates of elemental abundances
in the most metal-poor stars.  Although the number of lines suitable for
abundance analyses certainly decreases with lower abundances, the continuum
against which they are measured also gets cleaner, and uncertainties due to
contamination from blends lessens as well. We recall that McWilliam \etal
(1995) suggested that the observed spread at very low metallicities could be
due to metallicity dependent yields.  From the sets of yields analysed here we
can exclude this possibility, at least for the $\alpha$-elements and Fe, since
they depend only slightly on the initial stellar metallicity and, again,
the global metallicity increases quite fast in the early phases of Galactic
evolution as indicated by the absence of zero-metallicity stars in the Galaxy.

-Finally, we predicted that in spite of the adopted yields, the [O/Fe] ratio
at high redshifts should show a sharp rise and we suggested that the same
behaviour is expected for the [O/Zn] ratio which should be not affected
by dust. Future measurements of either [$\alpha$/Zn] or [$\alpha$/Fe] ratios
in very metal poor stars will be very useful to infer the nature and age of
high-redshift objects.

\acknowledgements
CC acknowledges that this work was partially supported by CNPq and FAPESP, Brazil.  
TCB acknowledges partial support of this work from NSF grant AST 95-29454 to Michigan State
University. TCB expresses gratitude to the Instituto Astron\^omico e Geof\' {\i}sico of the University of S\~ao Paulo (IAG/USP) for hospitality
during his visit, when the final version of this paper was completed and
acknowledges that this visit was supported by FAPESP and pr\'o-reitoria
de p\'os-gradua\c c\~ao of the University of S\~ao Paulo, Brazil. 
KN acknowledges that this research has been supported in part
by the Grant-in-Aid for Scientific Research (05242102, 06233101) and COE
research (07CE2002) of the Japanese Ministry of Education, Science, and
Culture.

\begin{table}
\caption{Observational Data}
\smallskip
\label{}
\begin{tabular}{lcccccccr}
\hline
 & & & & & & & & \\
Reference            & O & Mg & Ca & Si & S & C & N & Fe \\
\noalign{\hrule}
Gratton \etal 1996     & x &    &    &    &   &   &   & x \\
Pilachowski \etal 1996 &   &  x &  x &  x &   &   &   & x \\
Porto de Mello 1996     &   & x  & x  & x  &   & x &   & x \\
Ryan \etal 1996        &   & x  & x  & x  &   &   &   & x \\
McWilliam \etal 1995  &   & x  & x  & x  &   &   &   & x \\
Tomkin \etal 1995      &   &    &    &    &   & x &   & x \\
Anderson \& Edvardsson 1994& &   &    &    &   & x &   & x \\
Nissen \etal 1994      &   &  x &  x &    &   &   &   & x \\
Primas \etal 1994      &   & x  & x  & x  &   &   &   & x \\
Edvardsson \etal 1993  &   & x  & x  & x  &   &   &   & x  \\
Gratton \& Sneden 1991  &   &    & x  & x  &   &   &   & x \\
Molaro \& Bonifacio 1990&   & x  & x  & x  &   &   &   & x \\
Zhao \& Magain 1990     &   &    & x  &    &   &   &   & x \\
Magain 1989             &   & x  &    &    &   &   &   & x \\
Gratton \& Sneden 1988  &   & x  & x  & x  &   &   &   & x \\
Carbon \etal 1987      &   &    &    &    &   & x & x & x \\
Fran\c cois 1987, 1988  &   &    &    &    & x &   &   & x \\
Gratton \& Sneden 1987  &   & x  &    &    &   &   &   & x \\
Fran\c cois 1986     &   & x  &    & x  &   &   &   & x \\
Tomkin \etal 1986      &   &    &    &    &   & x &   & x \\
Tomkin \etal 1985      &   & x  & x  & x  &   &   &   & x \\
Laird 1985       &   &    &    &    &   & x & x & x \\
Clegg \etal 1981       &   &    &    &    & x & x & x & x \\
& & & & & & & &  \\
\hline
\end{tabular}
\end{table}

\small
\begin{table}
\caption{Dispersion Analysis}
\smallskip
\label{}
\begin{tabular}{ccccc}
\noalign{\hrule\medskip}
Ratio & $ \,\,\,\,\,\,\,\,[Fe/H] \geq -0.5$ &   $-1.5 < [Fe/H] \leq -0.5$ & $ -2

.5 < [Fe/H] \leq -1.5$ &  $ \,\,\,\,\,\,\,\,[Fe/H] \leq -2.5$ \cr
 & & & & \cr
\noalign{\hrule\medskip}
$S_{BI}$ [C/Fe] &  0.16 (N=202) & 0.19 (N=76) &         0.21  (N=83)  &        0
.33   (N=23)\cr
$IS_{68}$     &  (0.15-0.17) & (0.17-0.21) &          (0.19-0.23)  &         (0.
28-0.47) \cr
$R_{95}$ &  0.63 & 0.75 & 0.82 &        1.29 \cr
\noalign{\smallskip\hrule\smallskip}
$S_{BI}$ [Ca/Fe] & 0.06  (N=202)  &    0.08 (N=72) &          0.12  (N=67)  &   
     0.18   (N=46)\cr
$IS_{68}$      &  (0.06-0.07)   &    (0.07-0.08) &          (0.11-0.13)   &     
   (0.17-0.21) \cr
$R_{95}$ &  0.24          &     0.31       &          0.47          &        0.7
1 \cr
\noalign{\smallskip\hrule\smallskip}
$S_{BI}$ [Mg/Fe] & 0.11  (N=220)  &    0.11  (N=80)   &      0.13   (N=70)  &   
    0.17   (N=48)\cr
$IS_{68}$     &  (0.10-0.12)   &    (0.10-0.12)    &       (0.12-0.14)  &       
  (0.16-0.22)\cr
$R_{95}$ &  0.43         &      0.43         &        0.51         &         0.6
7\cr
\noalign{\smallskip\hrule\smallskip}
$S_{BI}$ [N/Fe]  & 0.21 (N=67) & 0.30 (N=55) & \multispan 2 {\hskip 1 truecm 0.2
8$\,$(N=52)} \cr
$IS_{68}$      & (0.20-.23) & (0.27-0.36) & \multispan 2 {\hskip 1 truecm (0.23-
0.34)} \cr
$R_{95}$ & 0.82 & 1.18 & \multispan 2 {\hskip 1 truecm 1.10} \cr
\noalign{\smallskip\hrule\smallskip}
$S_{BI}$ [O/Fe] & 0.12 (N=74) & 0.13  (N=45) & \multispan 2 {\hskip 1 truecm 0.1
1$\,$(N=23)} \cr
$IS_{68}$& (0.11-0.13) & (0.12-0.14) & \multispan 2 {\hskip 1 truecm (0.09-0.15)
} \cr
$R_{95}$ &  0.47 & 0.51         & \multispan 2 {\hskip 1 truecm 0.43} \cr
\noalign{\smallskip\hrule\smallskip}
$S_{BI}$ [S/Fe] & 0.17 (N=27) & \multispan 2 {\hskip 0.8 truecm 0.09$\,$(N=17)} 
&  \cr
$IS_{68}$ & (0.16-0.21) & \multispan 2 {\hskip 0.8 truecm (0.08-0.10)} &  \cr
$R_{95}$ & 0.67 & \multispan 2 {\hskip 0.8 truecm0.35} &  \cr
\noalign{\smallskip\hrule\smallskip}
$S_{BI}$ [Si/Fe] & 0.07 (N=221) & 0.12 (N=77) & 0.15 (N=56) & 0.33 (N=43) \cr
$IS_{68}$    &  (0.06-0.07) & (0.11-0.13) & (0.14-0.19) & (0.31-0.37) \cr
$R_{95}$ &  0.27 & 0.47 & 0.59 & 1.29 \cr
\noalign{\medskip\hrule}
\end{tabular}
\end{table}
\normalsize

\vfill\eject
\begin{table}
\caption{Solar Abundances by Mass}
\smallskip
\label{}
\begin{tabular}{cccc}
\hline
 & & & \\
Element & Model A (WW95) & Model B (TNH96) & Observations (AG89) \\
\hline
 & & & \\
$^{12}C$ & 1.78 ($-$3) & 1.31 ($-$3) & 3.03 ($-$3) \\
$^{16}O$ & 7.15 ($-$3) & 7.74 ($-$3) & 9.59 ($-$3) \\
$^{14}N$ & 1.38 ($-$3) & 1.38 ($-$3) & 1.11 ($-$3) \\
$Mg$ & 2.48 ($-$4) & 3.27 ($-$4) & 5.15 ($-$4) \\
$Si$ & 7.04 ($-$4) & 6.06 ($-$4) & 7.11 ($-$4) \\
$S$ & 3.07 ($-$4) & 2.33 ($-$4) & 4.18 ($-$4) \\
$Ca$ & 3.95 ($-$5) & 3.44 ($-$5) & 6.20 ($-$5) \\
$Fe$ & 1.37 ($-$3) & 1.24 ($-$3) & 1.27 ($-$3) \\
$Z$  & 1.41 ($-$2) & 1.39 ($-$2) & 1.7 ($-$2) \\
 & & & \\
\hline
\end{tabular}
\end{table}

\vfill\eject
\centerline{\bf Figure Captions}
\smallskip
\noindent
\par\noindent
Fig. 1 -- Abundance ratios as function of the metallicity for O, Mg, and Ca.
The left panels show a comparison between the data (see Table 1) and the
models (dashed-line: with WW95 yields; dot-dashed line: with TNH96 yields). The
right panels show a comparison between the summary lines (as described in
section 3 -- the solid line represents the midmean and the dotted lines
represent the lower and upper semi-midmean) and the model predictions. The
parameter r represents the number of data points (10 or 15) in the estimation
window and s refers to amount of smoothing (the fraction of data included at
each location) in the summary lines.
\par
\noindent
Fig. 2 -- The same as Figure 1, for Si, S, C, and N.
\par
\noindent
Fig. 3 -- The biweight estimator of scale, $S_{BI}$, for the [X/Fe] ratios (X =
C, Ca, Mg, N, O, S, and Si) at different [Fe/H] intervals (see Table 2).
\par
\noindent
Fig. 4 -- Ejected masses in the form of $^{12}$C, $^{16}$O, and $^{24}$Mg
as function of the initial mass of the progenitor star given by
different authors (solid line: WW95, dot-dashed line: TNH96, dotted line:
Woosley and Weaver 1986, adopted by MF89).
\par
\noindent
Fig. 5 -- The same as Figure 4, for Si, Fe, Ca, and S.
\par
\noindent
Fig. 6 -- Masses ejected in the form of various elements, weighted by the IMF,
for the more massive stars, plotted as functions of the initial mass of the
progenitor (upper panel: WW95 yields, lower panel: TNH96 yields).
\par
\noindent
Fig. 7 -- The same as Figure 6, for $^{14}$N, based on the WW95 (solid line)
and TNH96 (dot-dashed line) yields.
\par
\noindent
Fig. 8 -- [O/Fe] plotted as function of the redshift, for different
cosmologies, adopting the WW95 yields (where T$_U$ is the age of the Universe,
$z_f$ initial redshift of galaxies formation, $\Lambda$ is the cosmological
constant).
\par
\noindent
Fig. 9 -- The same as figure 8, but adopting the TNH96 yields. 

\vfill\eject
 
\begin{figure*}
\centerline{\psfig{figure=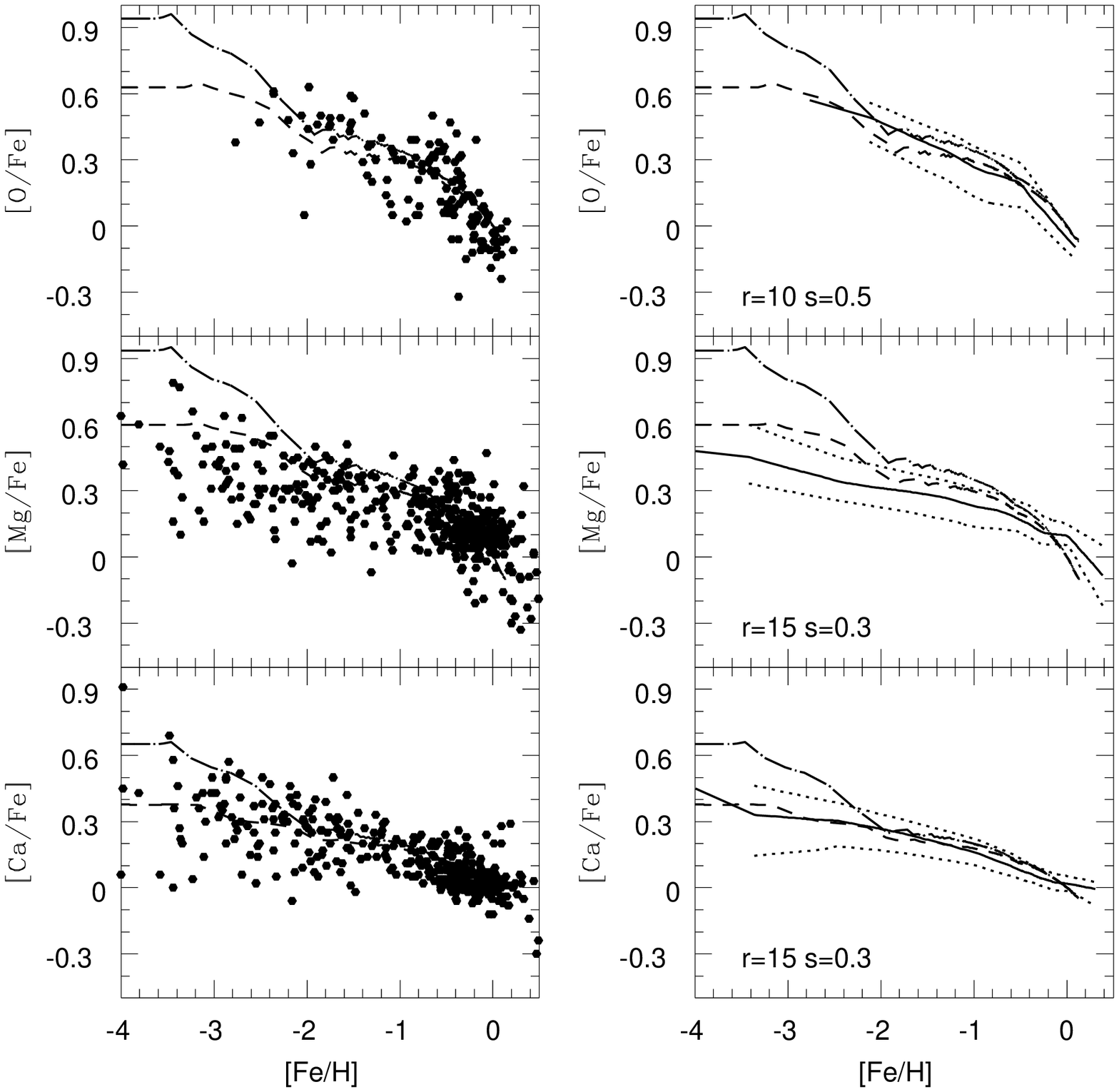,width=16cm,height=16cm} }
\caption{}
\end{figure*}
 
\clearpage
\eject
 
\begin{figure*}
\centerline{\psfig{figure=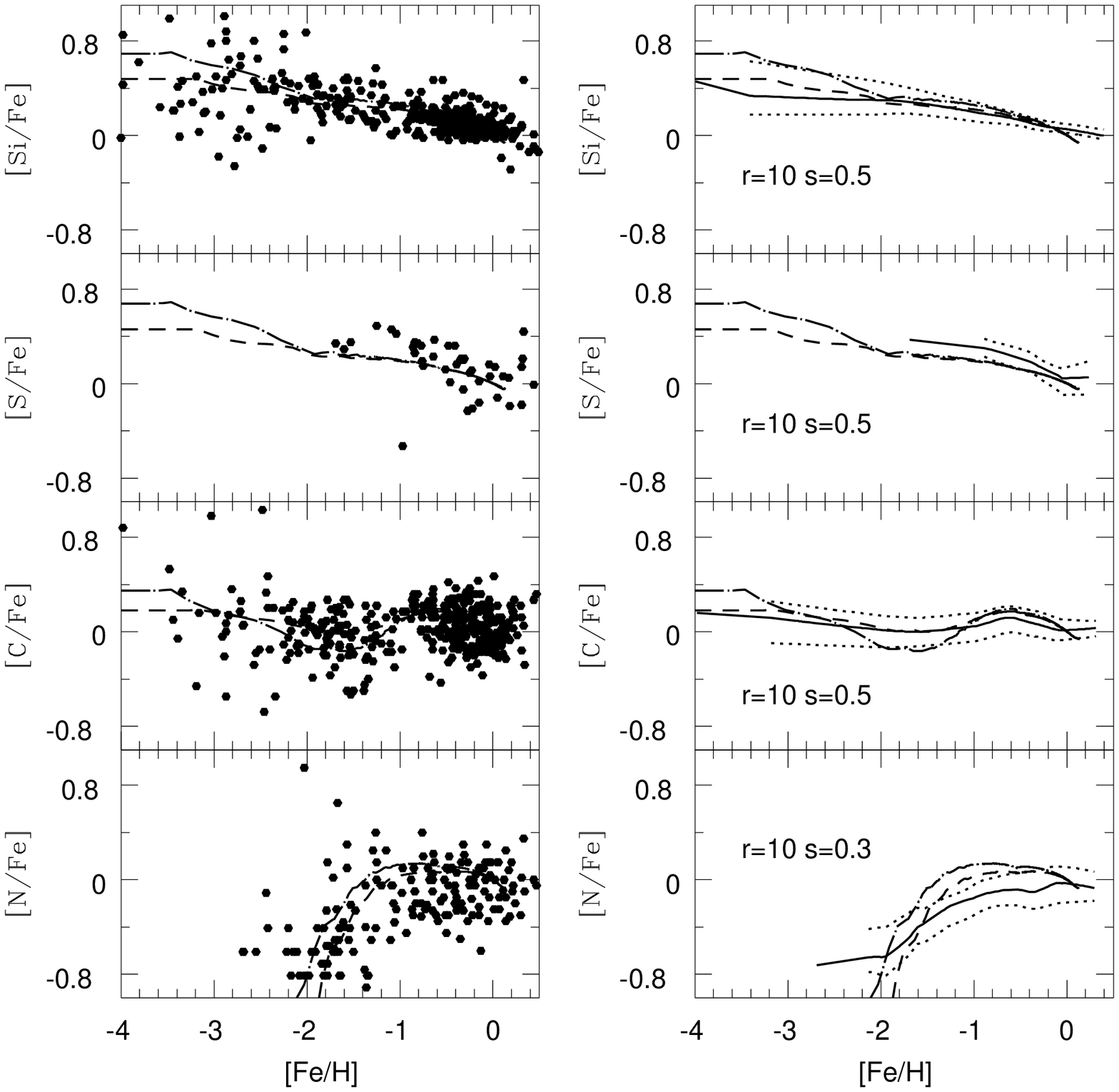,width=16cm,height=16cm} }
\caption{}
\end{figure*}
 
\clearpage
\eject
 
\begin{figure*}
\centerline{\psfig{figure=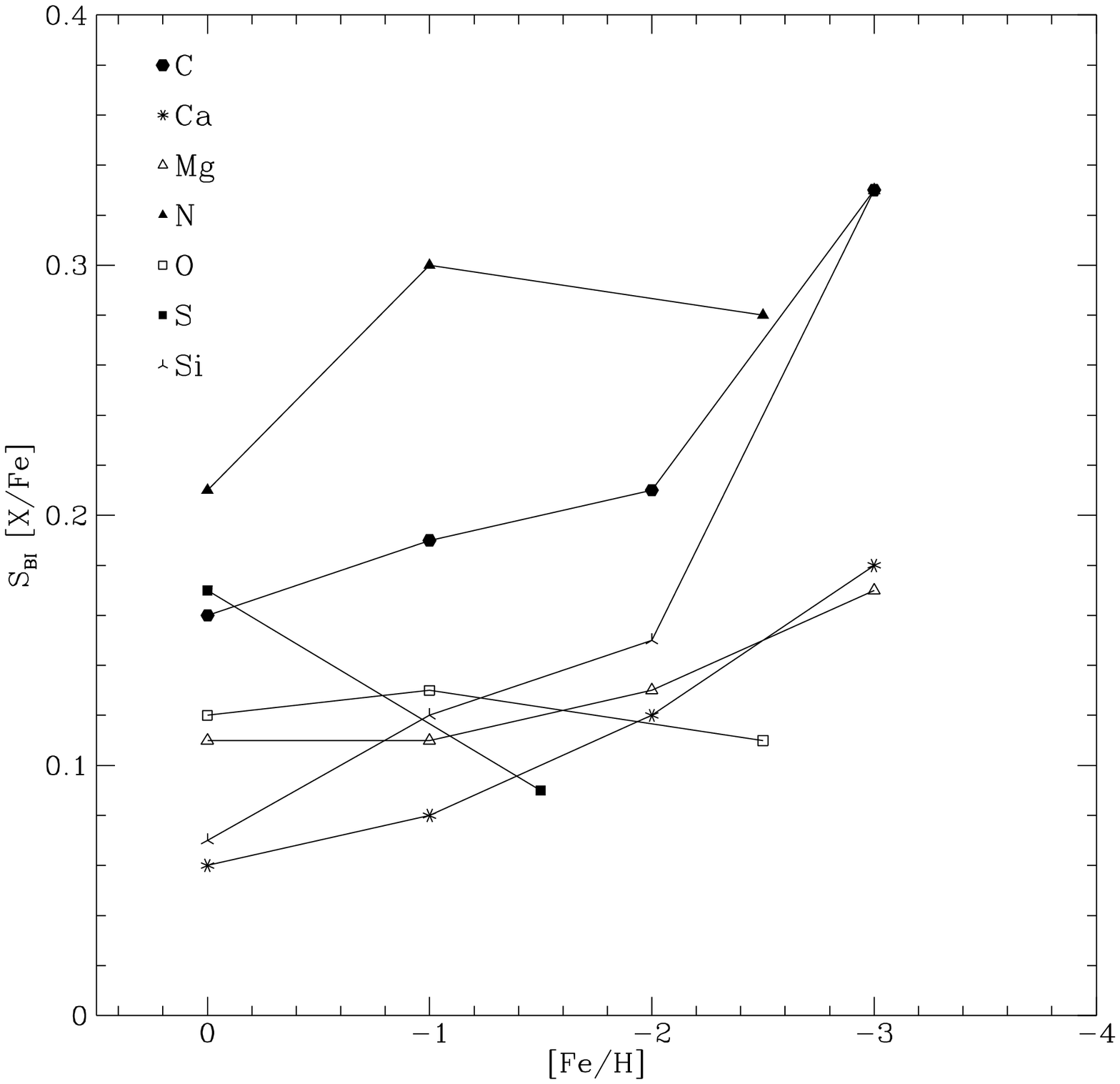,width=16cm,height=16cm} }
\caption{}
\end{figure*}
 
\begin{figure*}
\centerline{\psfig{figure=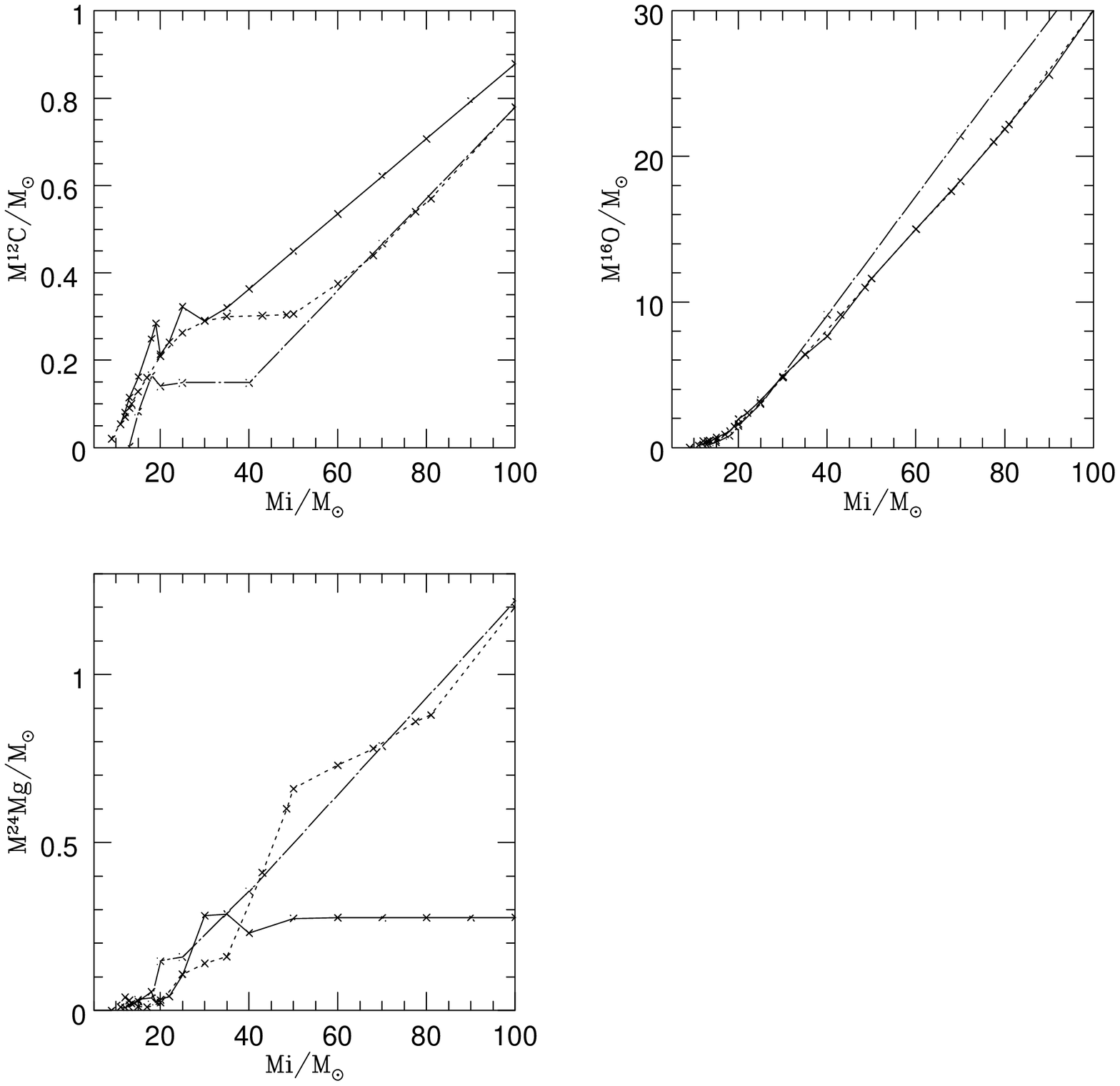,width=16cm,height=16cm} }
\caption{}
\end{figure*}
 
\begin{figure*}
\centerline{\psfig{figure=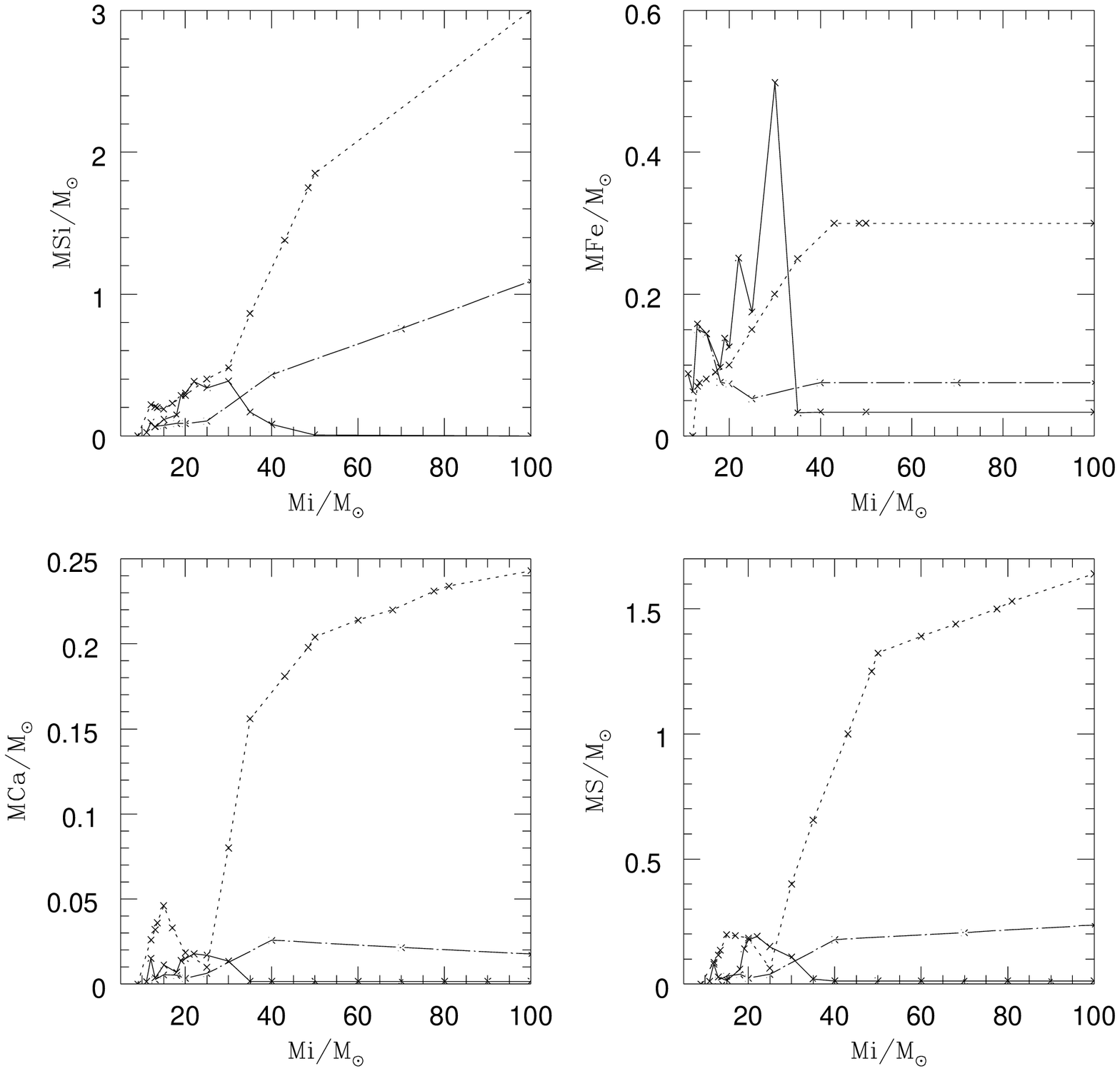,width=16cm,height=16cm} }
\caption{}
\end{figure*}
 
\begin{figure*}
\centerline{\psfig{figure=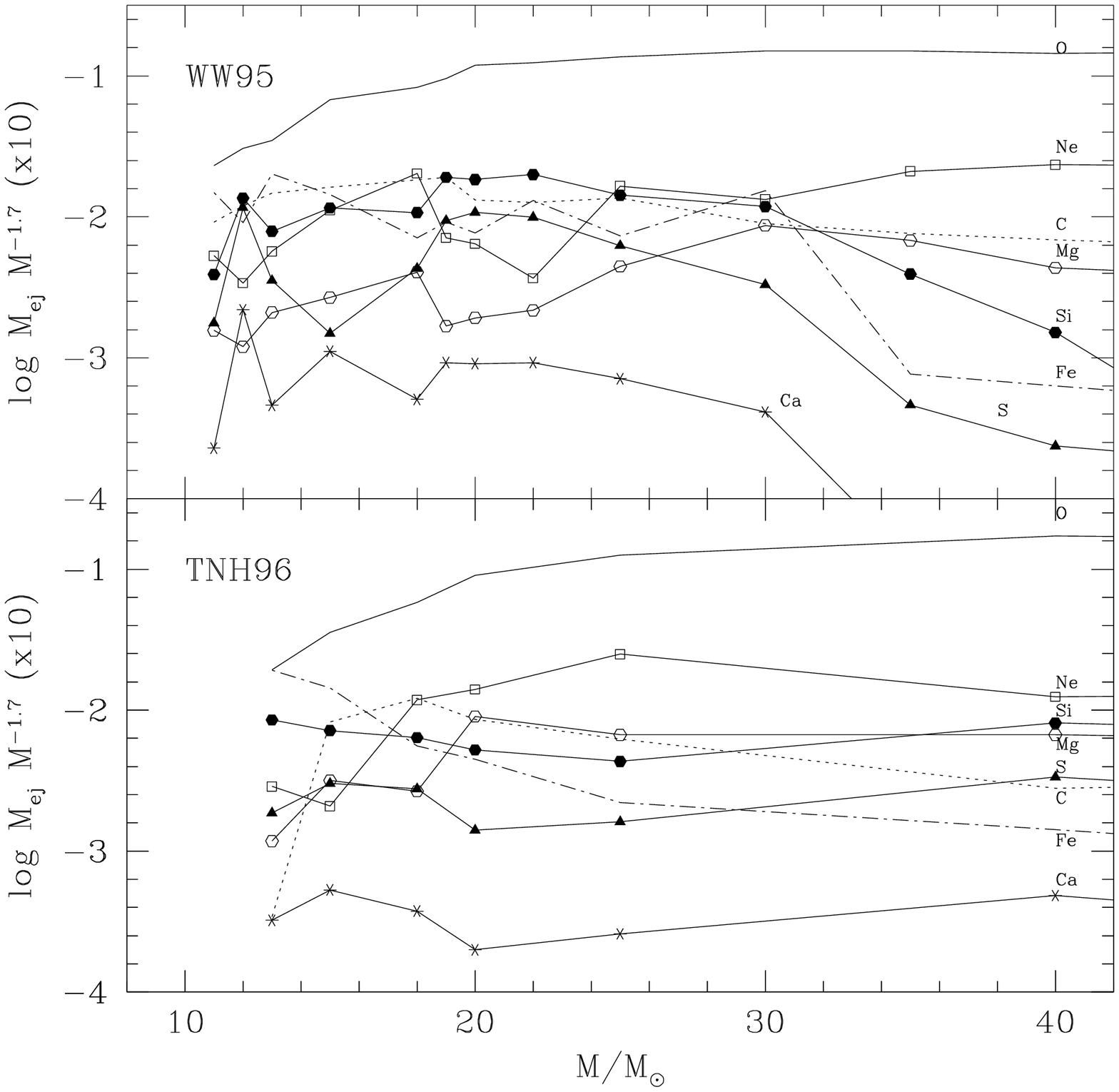,width=16cm,height=16cm} }
\caption{}
\end{figure*}

\begin{figure*}
\centerline{\psfig{figure=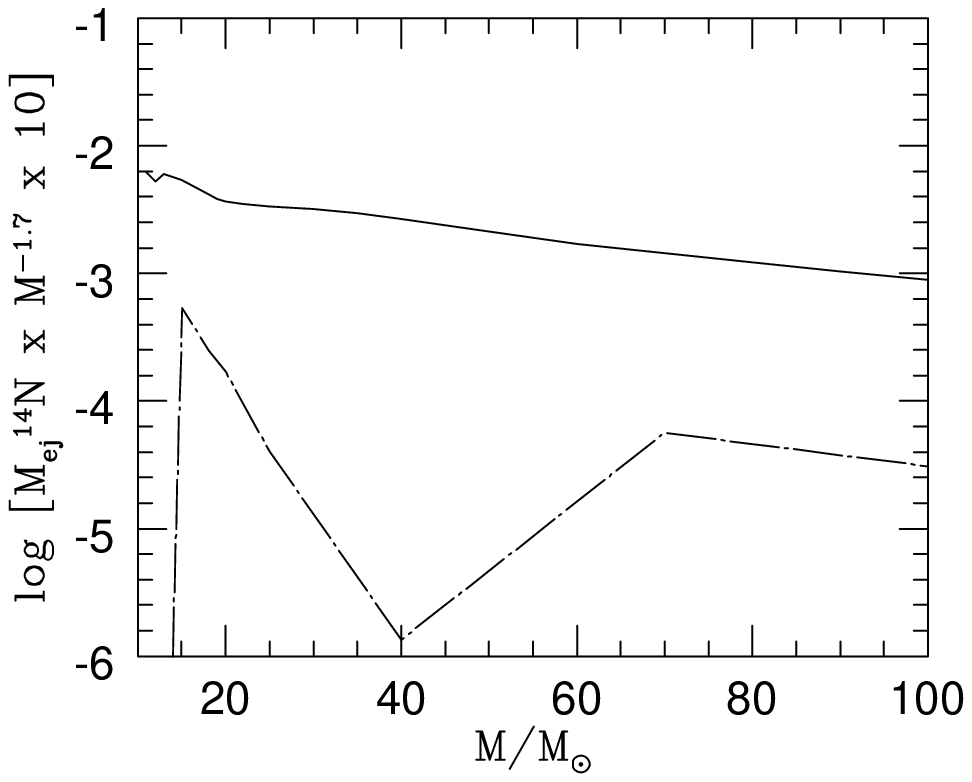,width=16cm,height=16cm} }
\caption{}
\end{figure*}
 
\begin{figure*}
\centerline{\psfig{figure=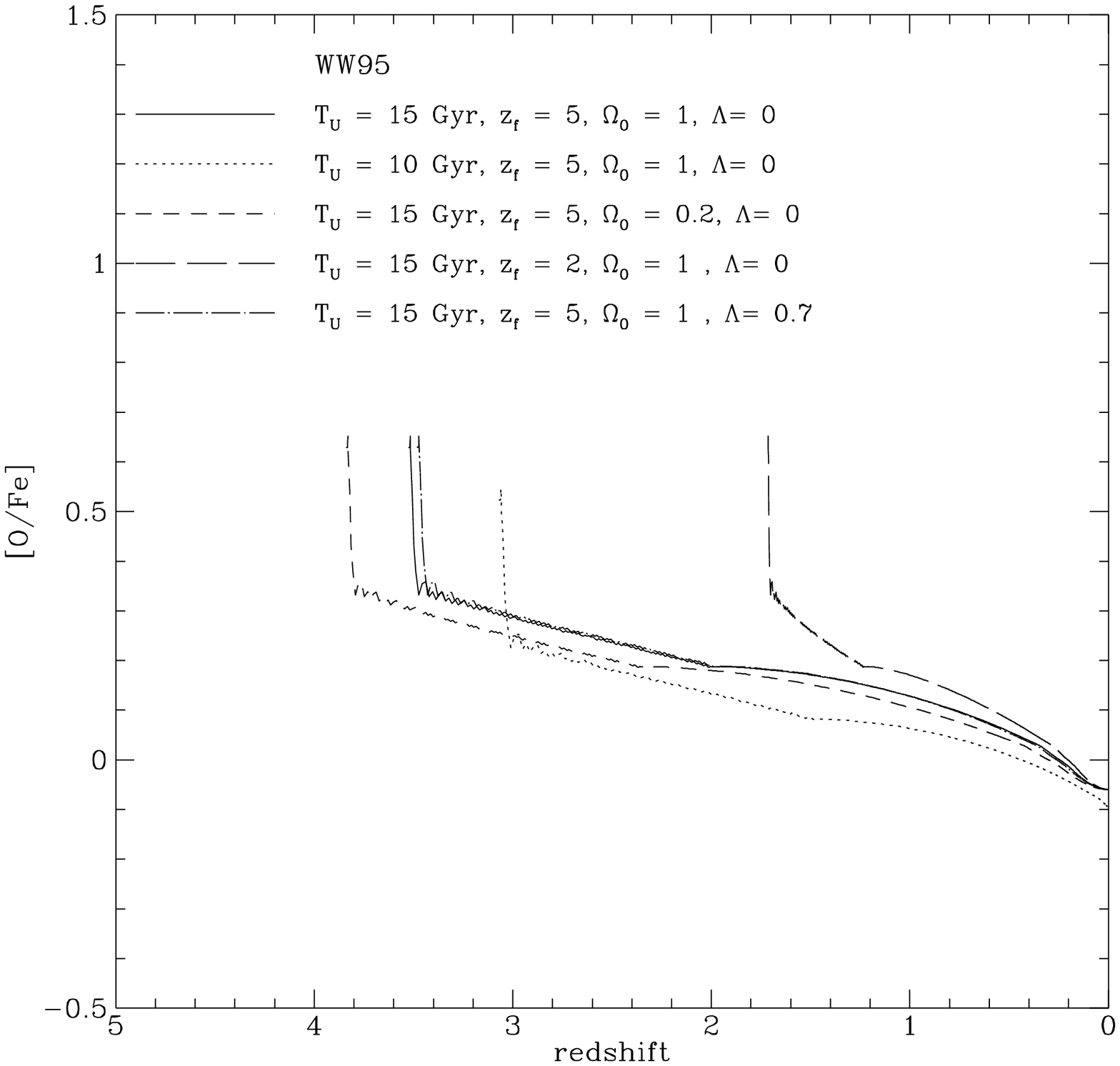,width=16cm,height=16cm} }
\caption{}
\end{figure*}

\begin{figure*}
\centerline{\psfig{figure=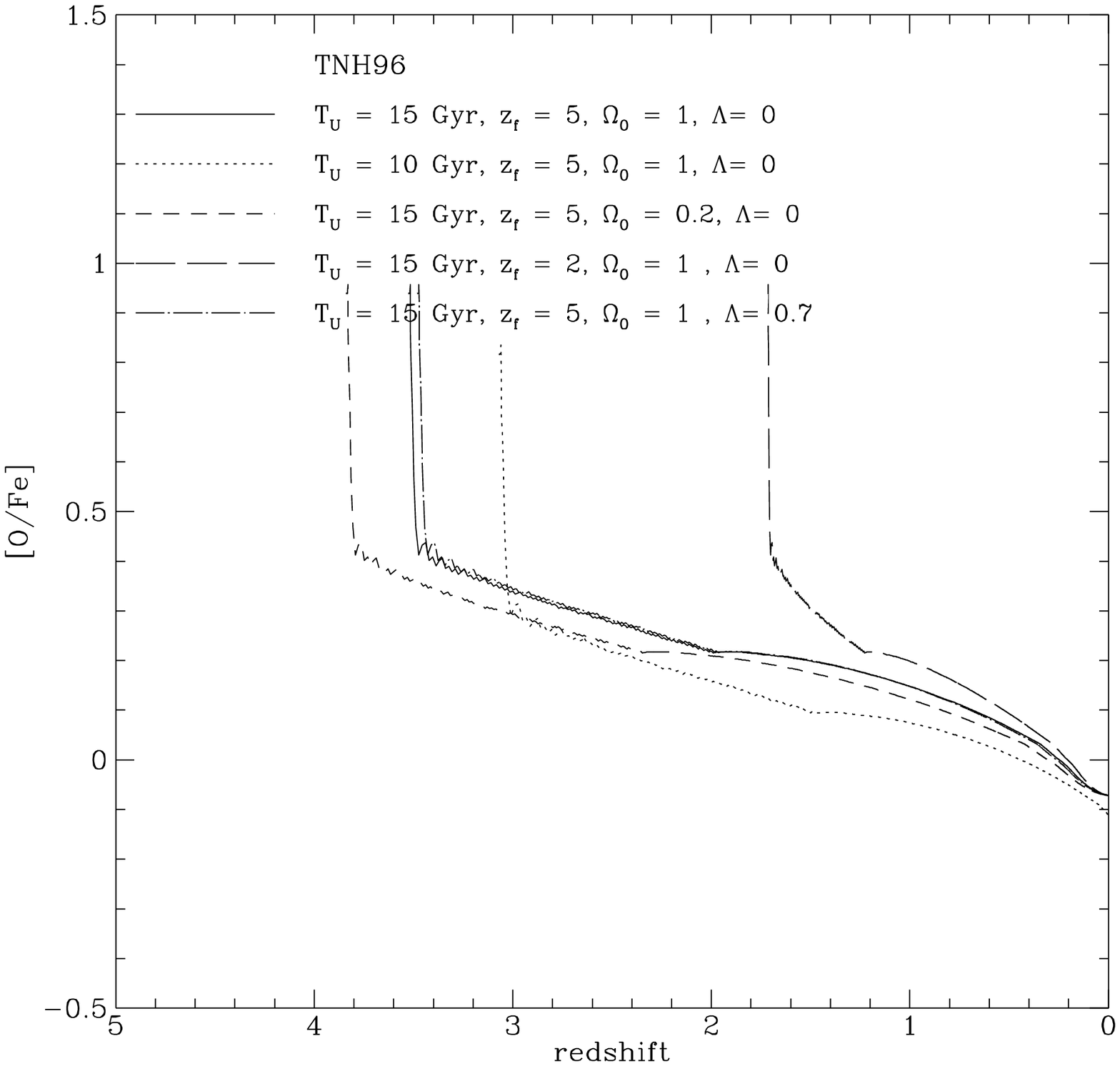,width=16cm,height=16cm} }
\caption{}
\end{figure*}

\end{document}